\documentstyle[12pt,twoside,fleqn,espcrc1,epsf]{article}

\newcommand{\nwc}{\newcommand}
\nwc{\cl}  {\clubsuit}
\nwc{\di}  {\diamondsuit}
\nwc{\sps} {\spadesuit}
\nwc{\hyp} {\hyphenation}
\nwc{\be}  {\begin{equation}}
\nwc{\ee}  {\end{equation}}
\nwc{\ba}  {\begin{array}}
\nwc{\ea}  {\end{array}}
\nwc{\bdm} {\begin{displaymath}}
\nwc{\edm} {\end{displaymath}}
\nwc{\bea} {\be\ba{rcl}}
\nwc{\eea} {\ea\ee}
\nwc{\ben} {\begin{eqnarray}}
\nwc{\een} {\end{eqnarray}}
\nwc{\bda} {\bdm\ba{lcl}}
\nwc{\eda} {\ea\edm}
\nwc{\bc}  {\begin{center}}
\nwc{\ec}  {\end{center}}
\nwc{\ds}  {\displaystyle}
\nwc{\bmat}{\left(\ba}
\nwc{\emat}{\ea\right)}
\nwc{\non} {\nonumber}
\nwc{\bib} {\bibitem}
\nwc{\lra} {\longrightarrow}
\nwc{\Llra}{\Longleftrightarrow}
\nwc{\ra}  {\rightarrow}
\nwc{\Ra}  {\Rightarrow}
\nwc{\lmt} {\longmapsto}
\nwc{\pa} {\partial}
\nwc{\iy}  {\infty}
\nwc{\ovl}  {\overline}
\nwc{\hm}  {\hspace{3mm}}
\nwc{\lf}  {\left}
\nwc{\ri}  {\right}
\nwc{\lm}  {\limits}
\nwc{\lb}  {\lbrack}
\nwc{\rb}  {\rbrack}
\nwc{\ov}  {\over}
\nwc{\pr}  {\prime}
\nwc{\nnn} {\nonumber \vspace{.2cm} \\ }
\nwc{\Sc}  {{\cal S}}
\nwc{\Lc}  {{\cal L}}
\nwc{\Rc}  {{\cal R}}
\nwc{\Dc}  {{\cal D}}
\nwc{\Oc}  {{\cal O}}
\nwc{\Cc}  {{\cal C}}
\nwc{\Pc}  {{\cal P}}
\nwc{\Mc}  {{\cal M}}
\nwc{\Ec}  {{\cal E}}
\nwc{\Fc}  {{\cal F}}
\nwc{\Hc}  {{\cal H}}
\nwc{\Kc}  {{\cal K}}
\nwc{\Xc}  {{\cal X}}
\nwc{\Gc}  {{\cal G}}
\nwc{\Zc}  {{\cal Z}}
\nwc{\Nc}  {{\cal N}}
\nwc{\fca} {{\cal f}}
\nwc{\xc}  {{\cal x}}
\nwc{\Ac}  {{\cal A}}
\nwc{\Bc}  {{\cal B}}
\nwc{\Uc}  {{\cal U}}
\nwc{\Vc}  {{\cal V}}
\nwc{\Th} {\Theta}
\nwc{\th} {\theta}
\nwc{\vth} {\vartheta}
\nwc{\eps}{\epsilon}
\nwc{\si} {\sigma}
\nwc{\Gm} {\Gamma}
\nwc{\gm} {\gamma}
\nwc{\bt} {\beta}
\nwc{\La} {\Lambda}
\nwc{\la} {\lambda}
\nwc{\om} {\omega}
\nwc{\Om} {\Omega}
\nwc{\dt} {\delta}
\nwc{\Si} {\Sigma}
\nwc{\Dt} {\Delta}
\nwc{\al} {\alpha}
\nwc{\vph}{\varphi}
\nwc{\zt} {\zeta}
\def\pr#1{#1^\prime}

\nwc{\Id}  {{\bf 1}}
\nwc{\diag} {{\rm diag}}
\nwc{\inv}  {{\rm inv}}
\nwc{\mod}  {{\rm mod}}
\nwc{\hal} {\frac{1}{2}}
\nwc{\tpi}  {2\pi i}
\def\KK{{\rm I\kern -.2em  K}}
\def\NN{{\rm I\kern -.16em N}}
\def\RR{{\rm I\kern -.2em  R}}
\def\ZZ{Z \kern -.43em Z}
\def\QQ{{\rm \kern .25em
             \vrule height1.4ex depth-.12ex width.06em\kern-.31em Q}}
\def\CC{{\rm \kern .25em
             \vrule height1.4ex depth-.12ex width.06em\kern-.31em C}}
\def\ZZZ{Z\kern -0.31em Z}

\def\pr#1{Phys. Rev. {\bf #1}}

\newcommand{\AmS}{{\protect\the\textfont2
  A\kern-.1667em\lower.5ex\hbox{M}\kern-.125emS}}

\title{QCD at high Baryon Density and Temperature:\\ 
Competing Condensates and the Tricritical Point}

\author{J{\"u}rgen Berges\address{Center for Theoretical Physics, \\ 
	Massachusetts Institute of Technology, Cambridge, MA 02139
\hfill MIT--CTP--2758}}

\begin{document}
% typeset front matter
\maketitle

\begin{abstract}
The phase diagram of strongly
interacting matter is explored as a function of temperature and
baryon number density. We investigate the
possible simultaneous formation of condensates in
the conventional quark--anti-quark channel 
(breaking chiral symmetry) and in a quark--quark channel
leading to color superconductivity: the spontaneous breaking of color
symmetry via the formation of quark Cooper pairs. We point out that 
for two massless quark flavors a
tricritical point in the phase diagram separates a chiral symmetry
restoring first order transition at high densities from the second order 
transition at high temperatures. Away from the chiral limit
this tricritical point becomes a second order phase transition 
with Ising model exponents, suggesting that a long correlation length
may develop in heavy ion collisions in which the
phase transition is traversed at the appropriate density.

\end{abstract}

\section{Towards the phase diagram}

\begin{figure}[t]
  \unitlength1.0cm
  \begin{center}
  \begin{picture}(8.0,3.5)
  \put(-1.5,-1.5){
  \epsfxsize=10.cm
  \epsfysize=5.cm
  \epsfbox[80 480 460 720]{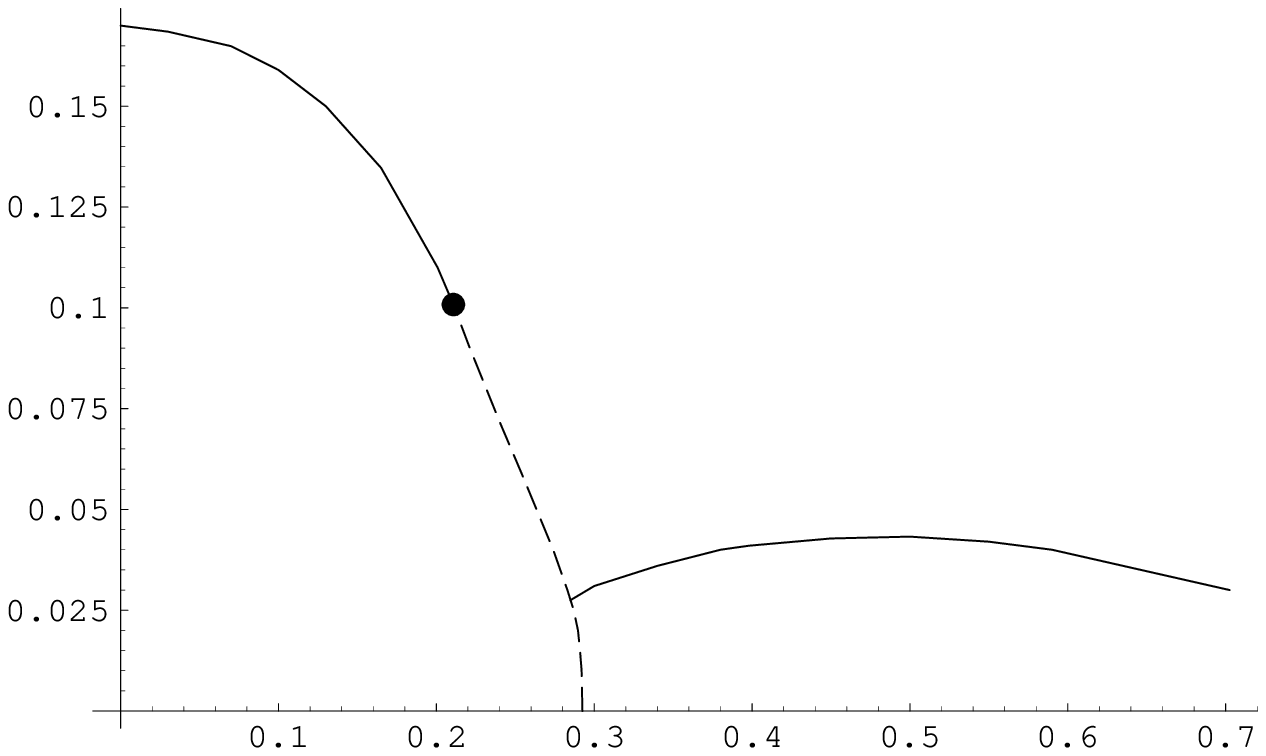}}
  \put(0.5,0.4){\bf $\small \langle\bar{\psi}\psi \rangle\not =0$}
  \put(3.8,-0.5){\bf $\small\langle\bar{\psi}\psi {\rangle}=0$}
  \put(4.8,2.8){\bf $\small\langle\bar{\psi}\psi {\rangle}=
  {\langle} \psi\psi \rangle=0$}
  \put(0.5,-0.3){\bf $\small{\langle} \psi\psi {\rangle}=0$}
  \put(5.8,-0.5){\bf $\small{\langle} \psi\psi {\rangle}\not =0$}
  \put(2.5,1.8){tricritical point}
  \put(-2.,1.4){\bf $T$}
  \put(9.,-0.9){\bf $\mu$} 
%  \put(2.5,4.8){\bf $\small \langle\bar{\psi}\psi \rangle\not =0$}
%  \put(7.8,1.5){\bf $\small\langle\bar{\psi}\psi {\rangle}=0$}
%  \put(7,5.3){\bf $\small\langle\bar{\psi}\psi {\rangle}=
%  {\langle} \psi\psi \rangle=0$}
%  \put(2.5,4.1){\bf $\small{\langle} \psi\psi {\rangle}=0$}
%  \put(7.8,0.8){\bf $\small{\langle} \psi\psi {\rangle}\not =0$}
%  \put(6.4,4.1){tricritical point}
%  \put(0.3,6.7){\bf $T$}
%  \put(12.8,0.3){\bf $\mu$} 
  \end{picture}
  \end{center}
%\vspace{-0.5cm}
\caption{Phase diagram as a function of $T$ and $\mu$ in GeV.  
The solid curves are second order phase transitions; the
dashed curve describes the first order transition. \label{phase}}
\end{figure}

The behavior of QCD at high temperature and baryon density is of 
fundamental interest and has applications in cosmology, in the
astrophysics of neutron stars and in the physics of heavy ion collisions.
Over the past years, considerable progress has been achieved in our 
understanding of high temperature QCD, where simulations on 
the lattice and universality arguments played an essential role. 
Recently, nonperturbative 
renormalization group methods have been established 
which account for both the low temperature chiral 
perturbation theory domain of validity and the domain of validity of 
universality associated with critical phenomena \cite{BJW}. They may help 
to shed some light on the remaining pressing questions at finite temperature, 
like the nature of the phase transition for realistic values of the quark 
masses. 

Our knowledge of the high density properties of strongly
interacting matter is rudimentary so far. There are severe problems
to use standard simulation algorithms at nonzero chemical potential 
on the lattice due to a complex fermion determinant. 
Different nonperturbative methods like the Exact Renormalization 
Group\footnote{First 
investigations in this direction done by D.\ Jungnickel, C.\ Wetterich
and myself \cite{BJW2} have been presented by D.\ Jungnickel, 
these proceedings.}  
or Schwinger--Dyson equations \cite{BlRo}
seem to present promising alternatives.
However, finding suitable nonperturbative approximation schemes 
often relies on some knowledge about propagators or the relevant
degrees of freedom. As a first step it seems well 
justified to consider models which allow us to describe likely
patterns of symmetry breaking and to make rough quantitative 
estimates. In this talk, I will present an exploration of the phase 
diagram for strongly interacting matter, using a class of models 
for two flavor QCD
in which the interaction between quarks is modelled
by that induced by instantons. The model is discussed using a mean field
approximation. The results are based on recent work \cite{BR} 
done in collaboration with Krishna Rajagopal.

QCD at high density is expected to have
a rich phase structure. In addition to the nuclear and quark matter phases a 
number of possibilities like the formation of meson condensates or 
strange quark matter have been discussed. A particularly interesting 
possibility is the formation of Cooper pairs of quarks, where 
an arbitrarily weak attractive 
interaction between quarks renders the quark Fermi surface unstable
and leads to the formation of a condensate. Since pairs of quarks
cannot be color singlets, a diquark condensate breaks color symmetry. 
For recent reviews of this
rapidly growing field of research see \cite{S,W}. Different 
attractive channels may lead to a (simultaneous) formation of different 
condensates. In particular, the vacuum of QCD already
has a condensate of a quark--anti-quark pair. An important ingredient
for the understanding of the high density phase structure is the
notion of {\it competing condensates}. 
One expects, and we find \cite{BR}, that the breaking of color symmetry 
due to a $\langle \psi \psi \rangle$ condensate is suppressed by the presence
of a chiral condensate. Likewise, we find chiral symmetry restoration 
to be induced at lower densities by the presence of a color superconductor
condensate. This behavior can be quantitatively understood from the 
effective potential which I discuss below.  
The competition between condensates may greatly simplify calculations.  
We find that previous treatments \cite{ARW,instanton}, which discuss the 
chiral and superconducting condensates separately and inspired this work, 
are a good approximation once the phase boundary is known. 
It is encouraging that this finding
has also recently been confirmed in calculations in an 
instanton liquid model \cite{CD}.   

The phase diagram which 
we uncover has striking qualitative features, several of which we 
expect to generalize beyond the model which we consider.
The phase diagram is shown in Fig. \ref{phase} for zero quark mass
as a function of temperature $T$ and chemical potential $\mu$ 
for (net) quark number. 
At low temperatures, we observe \cite{BR} chiral symmetry restoration via
a first order transition between a phase with low 
baryon density and a high density phase with a condensate
of quark--quark Cooper pairs in color antitriplet, Lorentz scalar,
isospin singlet states.\footnote{There are also indications\cite{ARW} of
a color ${\bf 6}$, Lorentz axial vector, isospin
singlet condensate which is many orders of magnitude
smaller than the condensates we treat.}
There are coexisting $\langle \psi \psi \rangle$
and $\langle\bar \psi \psi\rangle$ condensates in this phase
in the presence of a current quark mass.
We find color superconductivity 
at temperatures $T<T_c^\Delta$ where $T_c^\Delta$ is
of order tens to almost one hundred  
MeV. The transition we find at $T_c^\Delta$ is second order
for the present model, but the order of this transition 
may change once gauge field fluctuations are taken into
account.

In the chiral limit, we find a second order 
chiral transition at zero chemical potential, and a treatment of 
the present model which went beyond mean
field theory, e.g.\ along the lines presented in \cite{BJW}, would 
find critical 
exponents characteristic of the three dimensional $O(4)$ universality class. 
The point in the phase diagram where the second order transition
meets the low temperature first order transition locates
the {\it tricritical point}{}\footnote{Note that the presence of a 
tricritical point is 
precisely what is expected to occur at $\mu=0$ at a particular value of 
the strange quark mass \cite{wil,RW,GGP}. The two tricritical points are
continuously connected \cite{SRS}.}.  
The temperature is estimated in our model to be $T_{\rm tc}\simeq 101$ MeV 
and a chemical potential $\mu_{\rm tc}\simeq 211$ MeV.
This point has been observed
in different phenomenological models \cite{fir,klevansky,BR,stonyb}.
We note that the critical
behavior in the vicinity of this point is governed by universality 
(pointed out independently in \cite{stonyb}). If two-flavor QCD has a 
second order transition at
high temperatures and a first order transition at
high densities, then it will have a tricritical point
in the same universality class as our model.  
The physics near the tricritical point
is described by a $\phi^6$ field theory and there are three independent
critical exponents which are given quantitatively by our mean field
analysis and by logarithmic corrections to scaling \cite{tricrit}. 
A nonzero quark mass $m$ explicitly breaks
chiral symmetry and the second order phase transition above $T_{\rm tc}$
turns into a smooth crossover.
A small quark mass cannot eliminate the first order
transition below $T_{\rm tc}$. Therefore, whereas we previously had a line
of first order transitions and a line of second order
transitions meeting at a tricritical point, 
with $m\neq 0$ we now
have a line of first order transitions ending at an
ordinary critical point. The situation is precisely analogous to critical 
opalescence in a liquid gas system.
At this critical point, one degree of freedom (that associated
with the $\sigma$) becomes massless, while
the pion degrees of freedom are massive
since chiral symmetry is explicitly broken.  Therefore,
this transition is in the same universality class as
the three dimensional Ising model.   

From many studies of QCD at nonzero temperature, we are
familiar with the possibility of a second order transition,
with infinite correlation lengths, in an unphysical world
in which there are two massless quarks. It is exciting
to realize that if the finite density transition is first
order at zero temperature (see also \cite{BJW2}) 
then there is a tricritical
point in the chiral limit which becomes an Ising second
order phase transition in a world with chiral symmetry
explicitly broken. In a sufficiently energetic heavy ion collision, one 
may create conditions in approximate local thermal
equilibrium in the phase in which spontaneous chiral symmetry 
breaking is lost.
Depending on the initial density and temperature, when
this plasma expands and cools it will traverse the 
phase transition at different points in the ($\mu,T)$ plane.
Our results suggest that in
heavy ion collisions in which the chiral symmetry breaking
transition is traversed at baryon densities which
are not too high and not too low, 
a very long correlation length in the $\sigma$ channel
may be manifest even though the pion is massive \cite{BR,stonyb}.
Recently, a number of distinctive signatures have been proposed 
\cite{SRS} that could allow an identification of the predicted 
critical point.

\section{Effective potential} 

Our results are obtained from a class of models 
for QCD where the fermions interact via the
instanton induced interactions between light quarks. 
The interaction reflects the
chiral symmetry of QCD: axial baryon number is broken, while
chiral $SU(2)_L\times SU(2)_R$ is respected. Color $SU(3)$ is realized
as a global symmetry. We note that
with the help of appropriate Fierz transformations the instanton interaction 
can be decomposed into two parts, where
one part contains only color singlet fermion bilinears and the other part 
contains only color $\bar{\bf{3}}$ bilinears: 
$S_I=S_I^{(\bf{1}_c)}+S_I^{(\bar{\bf{3}}_c)}$ with (see \cite{BR})
\ben
S_{I}^{(\bf{1}_c)} &=& G_1 T \sum\limits_{n \in \ZZ}\int 
\frac{d^3 \vec{p}}{(2\pi)^3}
\left\{ 
-O_{(\sigma)}[\psi,\bar{\psi};-p] 
 O_{(\sigma)}[\psi,\bar{\psi};p]
-O^a_{(\pi)}[\psi,\bar{\psi};-p] 
 O_{(\pi)a}[\psi,\bar{\psi};p] \right.\nnn && \left. 
\qquad \qquad \qquad \,\,\,\,\,\,
+O_{(\eta^{\prime})}[\psi,\bar{\psi};-p] 
 O_{(\eta^{\prime})}[\psi,\bar{\psi};p]
+O^a_{(a_0)}[\psi,\bar{\psi};-p] 
 O_{(a_0)a}[\psi,\bar{\psi};p]\right\}
\, , \vspace{.3cm} 
\nonumber\\ 
S_{I}^{(\bar{\bf{3}}_c)} &=&  
G_2 T \sum\limits_{n \in \ZZ}\int \frac{d^3 \vec{p}}{(2\pi)^3} \left\{
-{O^{\dagger\alpha}_{(s)}}[\bar{\psi};p]
  O_{(s)\alpha}[\psi;p]
+{O^{\dagger\alpha}_{(p)}}[\bar{\psi};p]
  O_{(p)\alpha}[\psi;p] \right\}\ 
\label{si3}
\een
where we have generalized the interaction to 
allow $G_1$ and $G_2$ to take on independent values.
Here $p\equiv (2n\pi T,\vec{p})$ due to the bosonic nature
of the fermion bilinears.
The bilinears $O_{(\sigma)}$, $O^a_{(\pi)}$,
$O_{(\eta^{\prime})}$ and $O^a_{(a_0)}$ with $a=1,2,3$ carry the quantum 
numbers associated with the scalar isosinglet ($\sigma$), the 
pseudo-scalar isotriplet ($\pi$), the pseudo-scalar 
isosinglet ($\eta^{\prime}$)
and the scalar isotriplet ($a_0$), respectively. Similarly, 
the bilinear $O_{(s)}^{\alpha}$ ($O_{(p)}^{\alpha}$), 
with the color index $\alpha=1,2$ or $3$, 
carries the quantum numbers of the color 
antitriplet scalar (pseudo--scalar) diquark. 
For the $\sigma$ and for the scalar diquark they read
\begin{eqnarray}
O_{(\sigma)}[\psi,\bar{\psi};p] &=& -i T \sum\limits_{n \in \ZZ} \int
\frac{d^3 \vec{q}}{(2\pi)^3}
F(\vec{q}) F(\vec{p}-\vec{q}) {\bar{\psi}}^{i}_{\alpha}(-q) 
{\psi_i}^{\alpha}(p-q) \ , \nonumber \\
O_{(s)}^{\alpha}[\psi;p] &=& T \sum\limits_{n \in \ZZ} \int
\frac{d^3 \vec{q}}{(2\pi)^3}
F(-\vec{q}) F(-\vec{p}+\vec{q})\, {{(\psi^T)}^i}_{\beta}(-p+q)\, C \gamma^5\, 
\epsilon^{\alpha\beta\gamma}\epsilon_{ij}\,
{\psi^j}_\gamma(-q) \ ,
\label{osigma}
\end{eqnarray}
where $C$ denotes the charge conjugation matrix and we have 
supplemented the bilinears by suitable form factors  
$F(\vec{q})$ to mimic the effects of asymptotic freedom.
The results we quote in our exploration of
the phase diagram are obtained
using the smooth form factor 
$F(\vec{q})=\Lambda^2/({\vec{q}}^{\,2}+\Lambda^2)$
with $\Lambda=0.8$ GeV 
and with $G_1$ fixed by requiring
a constituent quark mass of $400$ MeV at $\mu=T=m=0$ in order to
obtain a reasonable, albeit qualitative, phenomenology
and $G_2=3G_1/4$ which is motivated in \cite{BR}. We note that the
qualitative features which we address do not depend
on this specific choice\footnote{There is a phenomenological upper bound 
which $G_2$ must satisfy. For $G_2 > 2 G_1$ color would be spontaneously 
broken in the vacuum.}.
   
We note from the signs in $S_I$ that if we choose 
the sign of $G_1$ such that the interaction in
the $\sigma$ channel is attractive, so that chiral symmetry
breaking is favored, we may expect condensation in
the $\pi$ and scalar diquark channels also.
In the chiral limit, one can always
make a rotation such that there is no $\pi$ condensate.
A condensate in the pseudoscalar diquark channel would break parity
spontaneously, but this seems not to be favored by the
interaction (\ref{si3}). We will use the model to explore condensates in
the $\sigma$ and scalar diquark channels, i.e.\ we consider possible 
simultaneous condensates,
${\big\langle}{O_{(\sigma)}}{\big\rangle} \sim 
{\big\langle} \overline{\psi} \psi{\big\rangle}$ for chiral symmetry
breaking and 
${\big\langle}{O_{(s)}^{\alpha}}{\big\rangle}\sim
{\big\langle} \psi \psi{\big\rangle}$ for breaking of color symmetry
$SU(3) \to SU(2)$.

The appropriate tool to study the phase structure of the model is
to consider the effective potential $\Omega$ (generating functional 
of 1PI Green functions at zero momentum). 
Using standard techniques \cite{BR} we compute the
mean field effective potential $\Omega(\phi,\Delta;\mu,T)$ as a function of
two `order parameters'\footnote{We note that the two flavor diquark condensate
is not a gauge invariant order parameter. In this respect it
is analogous to the electroweak sector of the standard model.
Likewise, there is the possibility of no sharp transition 
but rather a crossover
with qualitatively different physical behavior and the same
symmetries.}  $\phi$ and $\Delta$.
Extremizing $\Omega$ leads to coupled gap equations, whose solutions 
$\phi_0$ and $\Delta_0$ are related to the chiral condensate
and the condensate of Cooper pairs by 
$\phi_0 = 2 G_1 \Big{\langle}\bar{\psi}\psi \Big{\rangle}$, 
$\Delta_0 = 2 G_2 \Big{\langle} \psi\psi \Big{\rangle}$.
At its extrema it corresponds to the 
thermodynamic potential, related to the energy density $\epsilon$, 
the entropy density $s$, the quark number density $n$ and the pressure $P$ 
by
\be
\Omega(\phi_0,\Delta_0;\mu,T) = \epsilon -T s - \mu n = -P\ .
\ee
As the temperature, the chemical potential or the quark mass changes,
the effective potential can have several local minima in the 
$(\phi,\Delta)$ plane. 
Only the global minimum corresponds to the lowest free 
energy state and is favored. The analysis yields the phase structure 
presented in Fig.\ \ref{phase}.

\begin{figure}[t]
  \unitlength1.0cm
  \begin{center}
  \begin{picture}(13.0,4.7)
  \put(-2.0,0.0){
  \epsfxsize=8.2cm
  \epsfysize=5.cm
  \epsfbox[80 480 460 720]{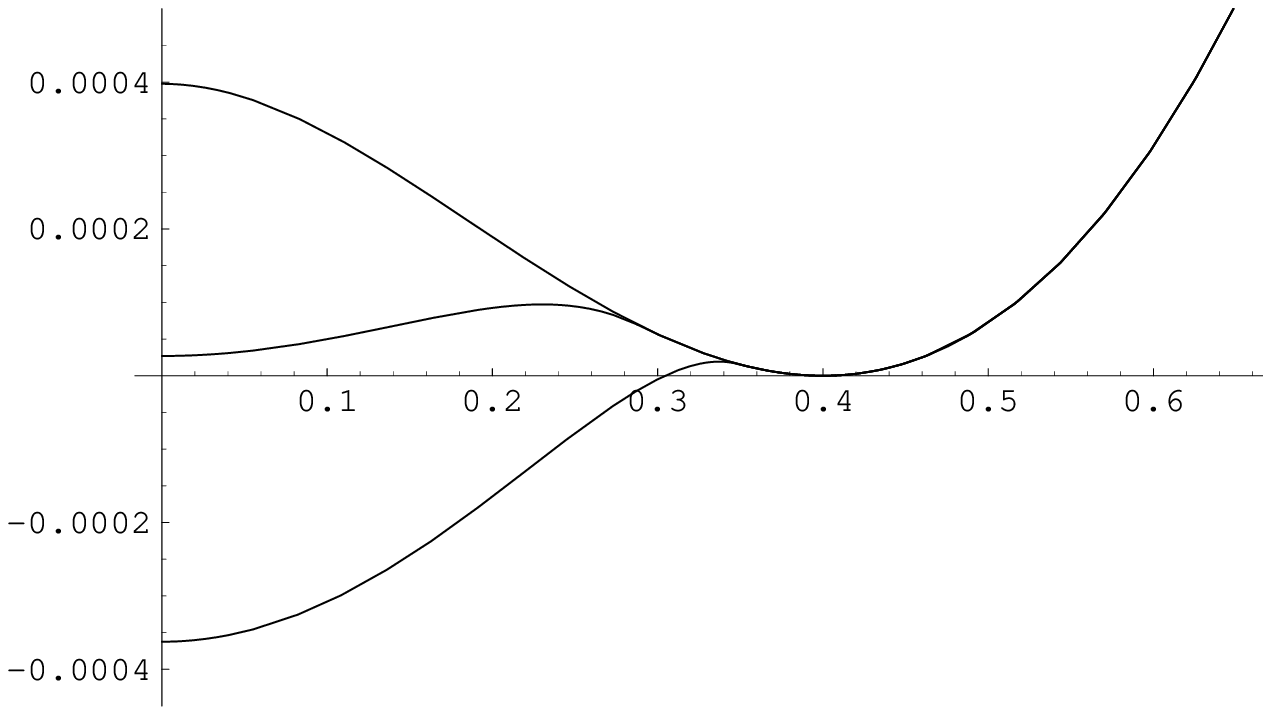}}
  \put(6.2,0.0){
  \epsfxsize=8.2cm
  \epsfysize=5.cm
  \epsfbox[80 480 460 720]{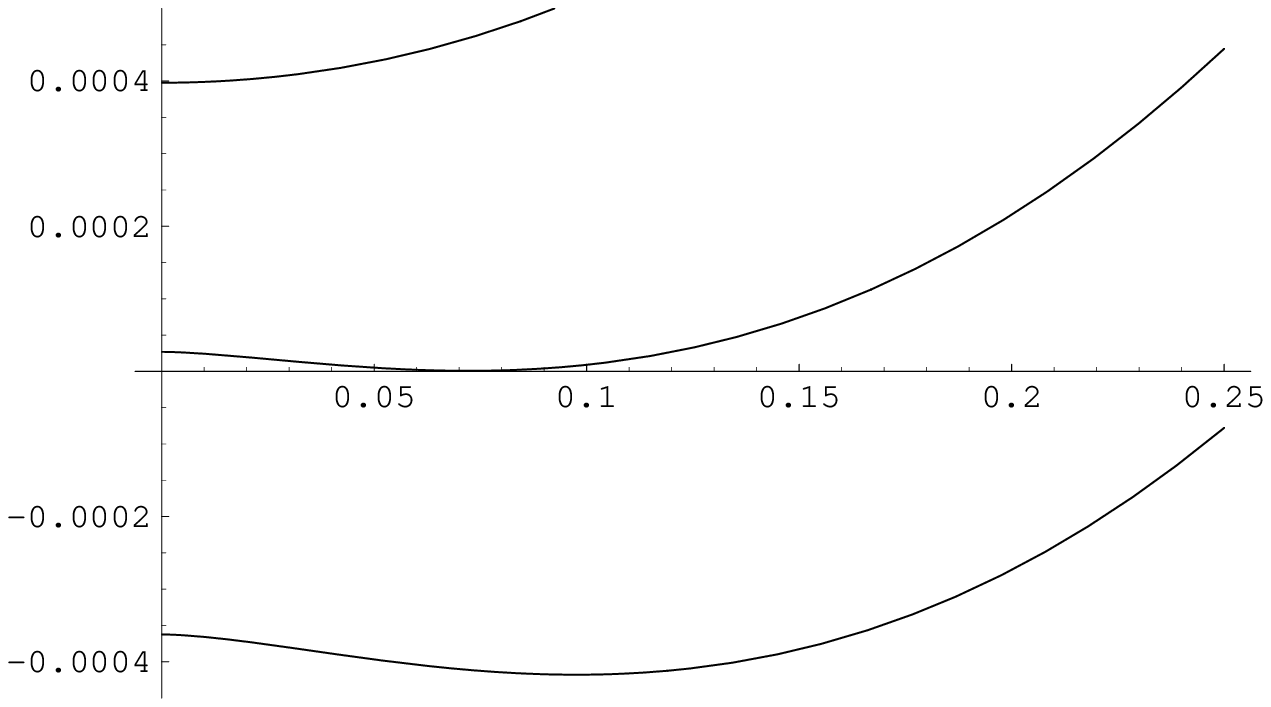}}
%  \put(-1.2,5.5){\bf $\Omega$}
%  \put(5.9,3.3){\bf $\phi$}
%  \put(7.0,5.5){\bf $\Omega$}
%  \put(14,3.3){\bf $\Delta$}
  \end{picture}
  \end{center}
%\centerline{
%\epsfysize=3.5in
%\hfill\epsfbox{twoslices.eps}\hfill
%}
\vspace*{-1.5cm}
\caption{The zero temperature thermodynamic potential 
$\Omega$ (in GeV$^4$) as a function
of $\phi \sim \Big{\langle}\bar{\psi}\psi \Big{\rangle}$ at $\Delta=0$ 
(left panel) and 
as a function of $\Delta \sim \Big{\langle} \psi\psi \Big{\rangle}$ at 
$\phi=0$ (right panel) for several chemical potentials. The curves 
correspond to (top to bottom) $\mu=0,0.292,0.35$ GeV.\label{pot}} 
%The curves at $\mu=\mu_0=0.292$ GeV are sections of Figure 1.}
\end{figure}
\begin{figure}[t]
  \unitlength1.0cm
  \begin{center}
  \begin{picture}(13.0,2.8)
  \put(-1.8,-1.4){
  \epsfxsize=8.2cm
  \epsfysize=4.5cm
  \epsfbox[80 480 460 720]{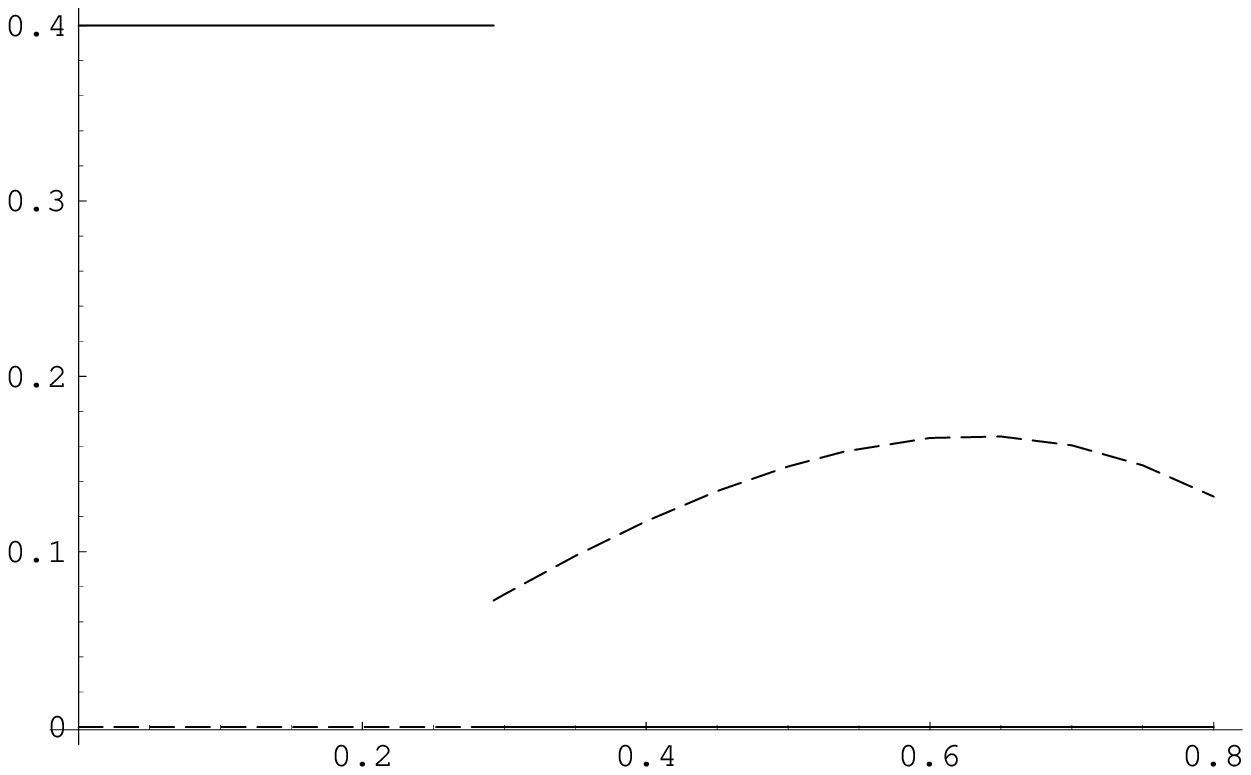}}
  \put(6.2,-1.4){
  \epsfxsize=8.2cm
  \epsfysize=4.5cm
  \epsfbox[80 480 460 720]{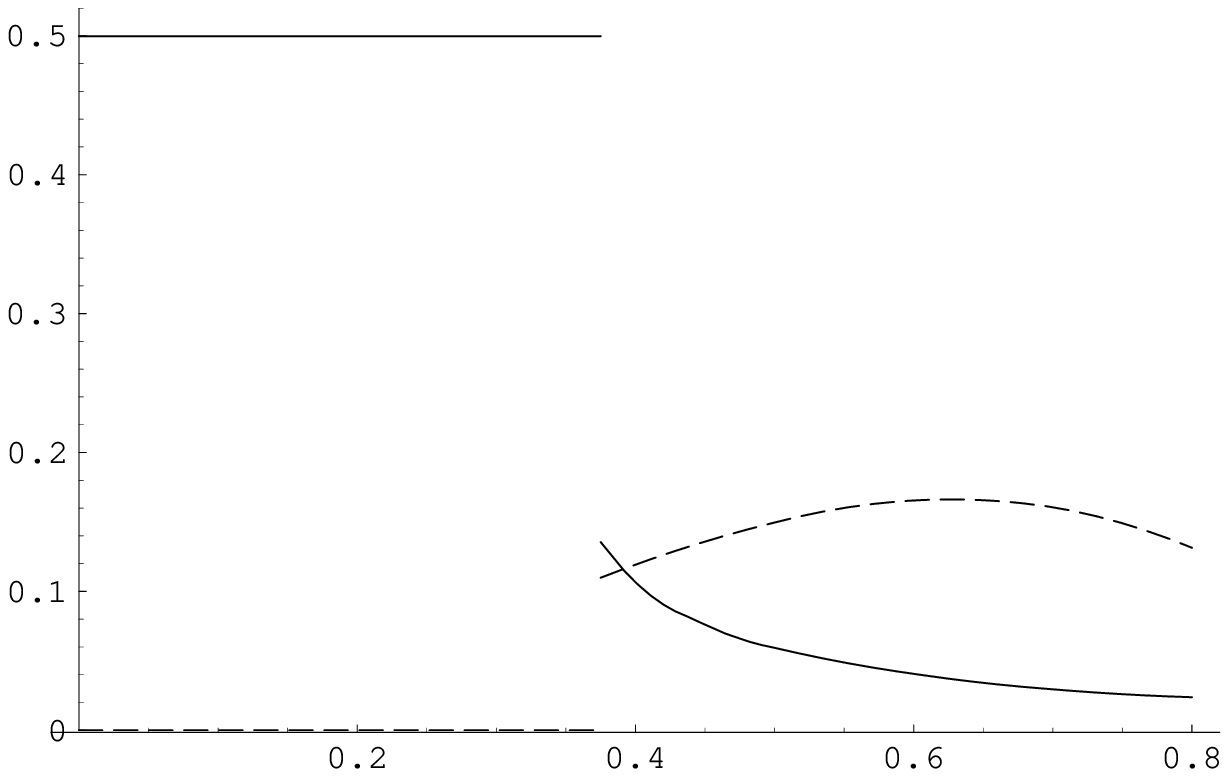}}
%  \put(7.5,4.9){\bf $n^{1/3}$}
%  \put(0,4.8){\bf $\phi_0$}
%  \put(3.7,2.5){\bf $\Delta_0$}
%  \put(2.5,0.2){\bf $\mu$}
%  \put(10.6,0.2){\bf $\mu$}
  \end{picture}
  \end{center}
%\vspace*{-1.8cm}
\caption{The chiral gap $\phi_0$ (solid line) and the diquark gap $\Delta_0$
(dashed line) as a function of the chemical potential $\mu$ in GeV 
for zero quark mass $m$
(left panel) and for $m=10$ MeV (right panel).\label{gap}}
\end{figure}

To discuss the competition between condensates let us concentrate on 
$T=0$. 
%As an example Fig.\ \ref{pot} shows $\Omega$
%at $T=0$ and $\mu=0.292$ GeV. One observes two degenerate minima 
%corresponding to a first order phase
%transition at which two phases have equal pressure and can coexist.
Fig.\ \ref{pot} shows `slices' of $\Omega$ as a function
of $\phi$ at $\Delta=0$ and 
as a function of $\Delta$ at $\phi=0$ 
for several chemical potentials. We find that  
$\Omega(\phi=0,\Delta)$ shows a nonzero minimum $\Delta_0$ for 
arbitrarily small $\mu$:
Since the instanton interaction provides an attractive interaction
in the color $\bar{3}$ scalar diquark channel it renders the Fermi
surface unstable and would lead to the formation of a diquark condensate.
However, one observes that the chiral condensate is favored for 
$\mu < 0.292$ GeV and the diquark condensate vanishes identically.
For larger $\mu$ chiral symmetry restoration occurs via a first
order transition, while the diquark condensate jumps to a nonzero 
value in the high density phase. We therefore have the situation 
that the competition with the chiral condensate leads to a first order
color superconductor transition which would be continuous otherwise.
One also notes from Figs.\ \ref{pot} that without considering the possibility 
of diquark condensation, i.e.\
for an identically vanishing $\Delta$, chiral symmetry restoration
would have occured at larger chemical potential when $\Omega(\phi,\Delta=0)$
crosses the origin. 
In the chiral limit we observe a {\it strong competition} in the sense 
that where one condensate is nonzero the other vanishes (see also Fig.\
\ref{gap}).

%\vspace*{-.5cm}
It is instructive to consider the physics away from the chiral limit.
A nonzero quark mass breaks chiral symmetry explicitly even in the high
density phase. Indeed, for an average current quark mass of $10$ MeV
one observes in Fig.\ \ref{gap} simultaneous 
condensates with comparable magnitude immediately upon the completion 
of the transition. Comparison with the chiral limit (left panel)
shows that the diquark condensate is not
significantly disturbed by the presence of $\phi_0$. This insensitivity 
can be understood by noting that although the Cooper pairs
have low momentum, they are formed from
quarks which have momenta close to the Fermi surface.
Adding a quark mass $m\ll \mu$ 
does not significantly affect the density of states 
or the interactions of the quasiparticles with momenta
of order $\mu$. 
%In fact, we observe a strong suppression of $\Delta_0$ as the constituent 
%quark mass $m+\phi_0$ becomes comparable to $\mu$. 
The phenomenon of color superconductivity
seems quite robust. Even away from the chiral limit it
is not significantly disturbed by the simultaneous 
presence of a chiral condensate in the high density phase 
as long as $m+\phi_0$ is smaller than the chemical potential.

%\vspace*{-.5cm}
\section{Outlook}
    
We have presented here a first 
model calculation of the QCD phase diagram that takes into account
the strong competition between the chiral and the color antitriplet scalar 
diquark condensate. The latter involves only the light $u$ and $d$ quarks 
of two colors. Though the strange quark is much heavier than the two 
light quarks, it may not be neglected for chemical potentials
much larger than $m_s$ or typical scales of a few hundred MeV.
There is a compelling symmetry breaking scheme \cite{ARW2}, involving
three massless flavors, which locks color and flavor to
a residual global $SU(3)$ symmetry. It is not clear what phase will be
realized with physical mass $u$, $d$ and $s$ quarks. As pointed out 
in \cite{S},
one may think of having the two flavor condensate first and color--flavor
locking at higher density. A first investigation may be performed
along the lines presented here \cite{BR} but finally more sophisticated
methods than the mean field analysis will be needed
to settle these questions. A promising possibility may be the use 
of truncated nonperturbative flow equations which have been successfully 
applied to finite temperature in the past.    
  
The phase diagram which we uncover in our two flavor study has striking
qualitative features, most remarkably the presence of a tricritical point.
For nonzero quark masses a situation similar to the liquid--gas
nuclear transition arises which has been much studied in low energy
heavy ion collisions. 
We observe the (tri)critical point to emerge as a result of 
a second order transition/crossover in one region 
of phase space (high $T$, small $\mu$) and a first order transition 
for the same order parameter in another region (high $\mu$, low $T$).  
The question of whether the tricritical point is 
realized in QCD is closely connected with the still unsettled question
of the order of the high temperature ($\mu=0$) transition, and
has to be addressed in a three flavor study. If the strange quark 
mass is too small, or if the axial $U(1)$ symmetry is effectively 
restored about the transition, then we may have a
first order transition at high $T$ which is driven by fluctuations 
\cite{PW}. In this case, a line of first order transitions connects the 
$T$ and $\mu$ axes.
However, one should note that if the finding \cite{KLM} of a 
continuous high density crossover for physical mass $u$, $d$ and $s$
quarks is realized in nature, then again a critical point in the phase
diagram may emerge: A high temperature first order transition line ending
in an Ising endpoint in the high density region. The involved long--range 
correlations in any of these scenarios would be appealing both from the 
theoretical and the experimental point of view.

%\vspace*{-0.1cm}

\end{document}